\documentclass[preprint,prc,aps,showpacs,groupedaddress,superscriptaddress,floatfix]{revtex4}

\input{tcilatex}

\begin{document}

\title{Any $l$-state approximate solutions of the Manning-Rosen potential by
the Nikiforov-Uvarov method}
\author{Sameer M. Ikhdair}
\email[E-mail: ]{sikhdair@neu.edu.tr}
\affiliation{Department of Physics, Near East University, Nicosia, Northern Cyprus, Turkey}
\author{Ramazan Sever}
\email[E-mail: ]{sever@metu.edu.tr}
\affiliation{Department of Physics, Middle East Technical University, 06531 Ankara,Turkey}
\date{\today}

\begin{abstract}
The Schr\"{o}dinger equation for the Manning-Rosen potential with the
centrifugal term is solved approximately to obtain bound states energies.
Additionally, the corresponding wave functions are expressed by the Jacobi
polynomials. The hypergeometric method (Nikiforov-Uvarov) (N-U) is used in
the calculations. To show the accuracy of our results, we calculate the
eigenvalues numerically for arbitrary quantum numbers $n$ and $l$ with two
different values of the potential parameter $\alpha .$ It is shown that the
results are in good agreement with the those obtained by other methods for
short potential range, small $l$ and $\alpha .$ This solution reduces to two
cases $l=0$ and Hulth\'{e}n potential case.

Keywords: Bound states; Manning-Rosen potential; Hypergeometric method.
\end{abstract}

\pacs{03.65.-w; 02.30.Gp; 03.65.Ge; 34.20.Cf}
\maketitle

\bigskip

\section{Introduction}

\noindent One of the important tasks of quantum mechanics is to find exact
solutions of the wave equations (nonrelativistic and relativistic) for
certain potentials of physical interest since they contain all the necessary
information regarding the quantum system under consideration. It is well
known that the exact solutions of these wave equations are only possible in
a few simple cases such as the Coulomb, the harmonic oscillator,
pseudoharmonic and Mie-type potentials [1-8]. For an arbitrary $l$-state,
most quantum systems could be only treated by approximation methods. For the
rotating Morse potential some semiclassical and/or numerical solutions have
been obtained by using Pekeris approximation [9-13]. In recent years, many
authors have studied the nonrelativistic and relativistic wave equations
with certain potentials for the $s$- and $l$-cases. The exact and
approximate solutions of these models have been obtained analytically
[10-14].

Many exponential-type potentials have been solved like the Morse potential
[12,13,15], the Hulth\'{e}n potential [16-19], the P\"{o}schl-Teller [20],
the Woods-Saxon potential [21-23], the Kratzer-type potentials
[12,14,24-27], the Rosen-Morse-type potentials [28,29], the Manning-Rosen
potential [29-33] and other multiparameter exponential-type potentials
[34,35]. Various methods are used to obtain the exact solutions of the wave
equations for this type of exponential potentials. These methods include the
supersymmetric (SUSY) and shape invariant method [19,36], the variational
[37], the path integral approach [31], the standard methods [32,33], the
asymptotic iteration method (AIM) [38], the exact quantization rule (EQR)
[13,39,40], the hypervirial perturbation [41], the shifted $1/N$ expansion
(SE) [42] and the modified shifted $1/N$ expansion (MSE) [43], series method
[44], smooth transformation [45], the algebraic approach [46], the
perturbative treatment [47,48] and the Nikiforov and Uvarov (N-U) method
[16,17,20--26,49-51] and others. The N-U method [51] is based on solving the
second-order linear differential equation by reducing to a generalized
equation of hypergeometric type. It has been used to solve the Schr\"{o}%
dinger [14,16,20,22,49], Dirac [50], Klein-Gordon [17,21,24,25] wave
equations for such kinds of exponential potentials.

Recently, the hypergeometric method (N-U) has shown its power in calculating
the exact energy levels of all bound states for some solvable quantum
systems. In this work, we attempt to apply this method to study another
exponential-type potential proposed by Manning and Rosen (M-R) [29-33]. With
an approximation to centrifugal term, we solve the Schr\"{o}dinger equation
to its bound states energies and wavefunctions. This potential is defined as
[29-33]%
\begin{equation}
V(r)=-\frac{A\hbar ^{2}}{2\mu b^{2}}\frac{e^{-r/b}}{1-e^{-r/b}}+\frac{\alpha
(\alpha -1)\hbar ^{2}}{2\mu b^{2}}\left( \frac{e^{-r/b}}{1-e^{-r/b}}\right)
^{2},
\end{equation}%
where $A$ and $\alpha $ are two-dimensionless parameters [27,28] but the
screening parameter $b$ has dimension of length which has a potential range $%
1/b.$ The potential (1) may be further put in the following simple form%
\begin{equation}
V(r)=-\frac{Ce^{-r/b}+De^{-2r/b}}{\left( 1-e^{-r/b}\right) ^{2}},\text{ }C=A,%
\text{ }D=-A-\alpha \text{(}\alpha \text{-}1)\text{,}
\end{equation}%
which is usually used for the description of diatomic molecular vibrations
[52,53]. It is also used in several branches of physics for their bound
states and scattering properties. The potential in (1) remains invariant by
mapping $\alpha \rightarrow 1-\alpha $ and has a relative minimum value $%
V(r_{0})=-\frac{A^{2}}{4\kappa b^{2}\alpha (\alpha -1)}$ at $r_{0}=b\ln %
\left[ 1+\frac{2\alpha (\alpha -1)}{A}\right] $ for $\alpha >0$ to be
obtained from the first derivative $\left. \frac{dV}{dr}\right\vert
_{r=r_{0}}=0.$ The second derivative which determines the force constants at 
$r=r_{0}$ is given by%
\begin{equation}
\left. \frac{d^{2}V}{dr^{2}}\right\vert _{r=r_{0}}=\frac{A^{2}\left[
A+2\alpha (\alpha -1)\right] ^{2}}{8b^{4}\alpha ^{3}(\alpha -1)^{3}}.
\end{equation}%
The contents of this paper are as follows: In Section II we breifly present
the hypergeometric method. In Section III, we derive $l\neq 0$ bound state
eigensolutions (eigenvalues and eigenfunctions) of the M-R potential by this
method. In Section IV, we present our numerical calculations for various
diatomic molecules. Section V, is devoted to for two special cases, namely, $%
l=0$ and the Hulth\'{e}n potential. The concluding remarks are given in
Section VI.

\section{\noindent The Hypergeometric Method}

The hypergeometric method (N-U) is based on solving the second-order linear
differential equation by reducing it to a generalized equation of
hypergeometric type [51]. In this method after employing an appropriate
coordinate transformation $z=z(r),$ the Schr\"{o}dinger equation can be
written in the following form:%
\begin{equation}
\psi _{n}^{\prime \prime }(z)+\frac{\widetilde{\tau }(z)}{\sigma (z)}\psi
_{n}^{\prime }(z)+\frac{\widetilde{\sigma }(z)}{\sigma ^{2}(z)}\psi
_{n}(z)=0,
\end{equation}%
where $\sigma (z)$ and $\widetilde{\sigma }(z)$ are the polynomials with at
most of second-degree, and $\widetilde{\tau }(s)$ is a first-degree
polynomial. The special orthogonal polynomials [51] reduce Eq. (4) to a
simple form by employing $\psi _{n}(z)=\phi _{n}(z)y_{n}(z),$ and choosing
an appropriate function $\phi _{n}(z).$ Consequently, Eq. (4) can be reduced
into an equation of the following hypergeometric type:%
\begin{equation}
\sigma (z)y_{n}^{\prime \prime }(z)+\tau (z)y_{n}^{\prime }(z)+\lambda
y_{n}(z)=0,
\end{equation}%
where $\tau (z)=\widetilde{\tau }(z)+2\pi (z)$ (its derivative must be
negative) and $\lambda $ is a constant given in the form%
\begin{equation}
\lambda =\lambda _{n}=-n\tau ^{\prime }(z)-\frac{n\left( n-1\right) }{2}%
\sigma ^{\prime \prime }(z),\text{\ \ \ }n=0,1,2,...
\end{equation}%
It is worthwhile to note that $\lambda $ or $\lambda _{n}$ are obtained from
a particular solution of the form $y(z)=y_{n}(z)$ which is a polynomial of
degree $n.$ Further, $\ y_{n}(z)$ is the hypergeometric-type function whose
polynomial solutions are given by Rodrigues relation%
\begin{equation}
y_{n}(z)=\frac{B_{n}}{\rho (z)}\frac{d^{n}}{dz^{n}}\left[ \sigma ^{n}(z)\rho
(z)\right] ,
\end{equation}%
where $B_{n}$ is the normalization constant and the weight function $\rho
(z) $ must satisfy the condition [51]%
\begin{equation}
w^{\prime }(z)-\left( \frac{\tau (z)}{\sigma (z)}\right) w(z)=0,\text{ }%
w(z)=\sigma (z)\rho (z).
\end{equation}%
In order to determine the weight function given in Eq. (8), we must obtain
the following polynomial:%
\begin{equation}
\pi (z)=\frac{\sigma ^{\prime }(z)-\widetilde{\tau }(z)}{2}\pm \sqrt{\left( 
\frac{\sigma ^{\prime }(z)-\widetilde{\tau }(z)}{2}\right) ^{2}-\widetilde{%
\sigma }(z)+k\sigma (z)}.
\end{equation}%
In principle, the expression under the square root sign in Eq. (9) can be
arranged as the square of a polynomial. This is possible only if its
discriminant is zero. In this case, an equation for $k$ is obtained. After
solving this equation, the obtained values of $k$ are included in the
hypergeometric method (N-U) and here there is a relationship between $%
\lambda $ and $k$ by $k=\lambda -\pi ^{\prime }(z).$ After this point an
appropriate $\phi _{n}(z)$ can be calculated as the solution of the
differential equation:%
\begin{equation}
\phi ^{\prime }(z)-\left( \frac{\pi (z)}{\sigma (z)}\right) \phi (z)=0.
\end{equation}

\section{Bound-state solutions for arbitrary $l$-state}

To study any quantum physical system characterized by the empirical
potential given in Eq. (1), we solve the original $\mathrm{SE}$ which is
given in the well known textbooks [1,2]

\begin{equation}
\left( \frac{p^{2}}{2m}+V(r)\right) \psi (\mathbf{r,}\theta ,\phi )=E\psi (%
\mathbf{r,}\theta ,\phi ),
\end{equation}%
where the potential $V(r)$ is taken as the M-R form in (1). Using the
separation method with the wavefunction $\psi (\mathbf{r,}\theta ,\phi
)=r^{-1}R(r)Y_{lm}(\theta ,\phi ),$ we obtain the following radial Schr\"{o}%
dinger eqauation as%
\begin{equation}
\frac{d^{2}R_{nl}(r)}{dr^{2}}+\left\{ \frac{2\mu E_{nl}}{\hbar ^{2}}-\frac{1%
}{b^{2}}\left[ \frac{\alpha (\alpha -1)e^{-2r/b}}{\left( 1-e^{-r/b}\right)
^{2}}-\frac{Ae^{-r/b}}{1-e^{-r/b}}\right] -\frac{l(l+1)}{r^{2}}\right\}
R_{nl}(r)=0,
\end{equation}%
Since the Schr\"{o}dinger equation with above M-R effective potential has no
analytical solution for $l\neq 0$ states$,$ an approximation to the
centrifugal term has to be made. The good approximation for $1/r^{2}$ in the
centrifugal barrier is taken as [18,33]%
\begin{equation}
\frac{1}{r^{2}}\approx \frac{1}{b^{2}}\frac{e^{-r/b}}{\left(
1-e^{-r/b}\right) ^{2}},
\end{equation}%
in a short potential range. To solve it by the present method, we need to
recast Eq. (12) with Eq. (13) into the form of Eq. (4) changing the
variables $r\rightarrow z$ through the mapping function $r=f(z)$ and energy
transformation given by%
\begin{equation}
z=e^{-r/b},\text{ }\varepsilon =\sqrt{-\frac{2\mu b^{2}E_{nl}}{\hbar ^{2}}},%
\text{ }E_{nl}<0,
\end{equation}%
to obtain the following hypergeometric equation:%
\[
\frac{d^{2}R(z)}{dz^{2}}+\frac{(1-z)}{z(1-z)}\frac{dR(z)}{dz} 
\]%
\begin{equation}
+\frac{1}{\left[ z(1-z)\right] ^{2}}\left\{ -\varepsilon ^{2}+\left[
A+2\varepsilon ^{2}-l(l+1)\right] z-\left[ A+\varepsilon ^{2}+\alpha (\alpha
-1)\right] z^{2}\right\} R(z)=0.
\end{equation}%
We notice that for bound state (real) solutions, the last equation requires
that%
\begin{equation}
z=\left\{ 
\begin{array}{ccc}
0, & \text{when} & r\rightarrow \infty , \\ 
1, & \text{when} & r\rightarrow 0,%
\end{array}%
\right.
\end{equation}%
and thus the finite radial wavefunctions $R_{nl}(z)\rightarrow 0.$ To apply
the hypergeometric method (N-U), it is necessary to compare Eq. (15) with
Eq. (4). Subsequently, the following value for the parameters in Eq. (4) are
obtained as 
\begin{equation}
\widetilde{\tau }(z)=1-z,\text{\ }\sigma (z)=z-z^{2},\text{\ }\widetilde{%
\sigma }(z)=-\left[ A+\varepsilon ^{2}+\alpha (\alpha -1)\right] z^{2}+\left[
A+2\varepsilon ^{2}-l(l+1)\right] z-\varepsilon ^{2}.
\end{equation}%
If one inserts these values of parameters into Eq. (9), with $\sigma
^{\prime }(z)=1-2z,$ the following linear function is achieved%
\begin{equation}
\pi (z)=-\frac{z}{2}\pm \frac{1}{2}\sqrt{a_{1}z^{2}+a_{2}z+a_{3}},
\end{equation}%
where $a_{1}=1+4\left[ A+\varepsilon ^{2}+\alpha (\alpha -1)\right] -k,$ $%
a_{2}=4\left\{ k-\left[ A+2\varepsilon ^{2}-l(l+1)\right] \right\} $ and $%
a_{3}=4\varepsilon ^{2}.$ According to this method the expression in the
square root has to be set equal to zero, that is, $\Delta
=a_{1}z^{2}+a_{2}z+a_{3}=0.$ Thus the constant $k$ can be determined as%
\begin{equation}
k=A-l(l+1)\pm a\varepsilon ,\text{ \ }a=\sqrt{(1-2\alpha )^{2}+4l(l+1)}.
\end{equation}%
In view of that, we can find four possible functions for $\pi (z)$ as%
\begin{equation}
\pi (z)=-\frac{z}{2}\pm \left\{ 
\begin{array}{c}
\varepsilon -\left( \varepsilon -\frac{a}{2}\right) z,\text{ \ \ \ for \ \ }%
k=A-l(l+1)+a\varepsilon , \\ 
\varepsilon -\left( \varepsilon +\frac{a}{2}\right) z;\text{ \ \ \ for \ \ }%
k=A-l(l+1)-a\varepsilon .%
\end{array}%
\right.
\end{equation}%
We must select%
\begin{equation}
\text{\ }k=A-l(l+1)-a\varepsilon ,\text{ }\pi (z)=-\frac{z}{2}+\varepsilon
-\left( \varepsilon +\frac{a}{2}\right) z,
\end{equation}%
in order to obtain the polynomial, $\tau (z)=\widetilde{\tau }(z)+2\pi (z)$
having negative derivative as%
\begin{equation}
\tau (z)=1+2\varepsilon -\left( 2+2\varepsilon +a\right) z,\text{ }\tau
^{\prime }(z)=-(2+2\varepsilon +a).
\end{equation}%
We can also write the values of $\lambda =k+\pi ^{\prime }(z)$ and $\lambda
_{n}=-n\tau ^{\prime }(z)-\frac{n\left( n-1\right) }{2}\sigma ^{\prime
\prime }(z),$\ $n=0,1,2,...$ as%
\begin{equation}
\lambda =A-l(l+1)-(1+a)\left[ \frac{1}{2}+\varepsilon \right] ,
\end{equation}%
\begin{equation}
\lambda _{n}=n(1+n+a+2\varepsilon ),\text{ }n=0,1,2,...
\end{equation}%
respectively. Additionally, using the definition of $\lambda =\lambda _{n}$
and solving the resulting equation for $\varepsilon ,$ allows one to obtain%
\begin{equation}
\varepsilon =\frac{(n+1)^{2}+l(l+1)+(2n+1)\Lambda -A}{2(n+1+\Lambda )},\text{
}\Lambda =\frac{-1+a}{2},
\end{equation}%
from which we obtain the discrete energy levels%
\begin{equation}
E_{nl}=-\frac{\hbar ^{2}}{2\mu b^{2}}\left[ \frac{(n+1)^{2}+l(l+1)+(2n+1)%
\Lambda -A}{2(n+1+\Lambda )}\right] ^{2},\text{ \ }0\leq n,l<\infty
\end{equation}%
where $n$ denotes the radial quantum number. It is found that $\Lambda $
remains invariant by mapping $\alpha \rightarrow 1-\alpha ,$ so do the bound
state energies $E_{nl}.$ An important quantity of interest for the M-R
potential is the critical coupling constant $A_{c},$ which is that value of $%
A$ for which the binding energy of the level in question becomes zero. Using
Eq. (26), in atomic units $\hbar ^{2}=\mu =Z=e=1,$%
\begin{equation}
A_{c}=(n+1+\Lambda )^{2}-\Lambda (\Lambda +1)+l(l+1).
\end{equation}

Let us now find the corresponding radial part of the wave function. Using $%
\sigma (z)$ and $\pi (z)$ in Eqs (17) and (21), we obtain%
\begin{equation}
\phi (z)=z^{\varepsilon }(1-z)^{(\Lambda +1)},
\end{equation}%
\begin{equation}
\rho (z)=z^{2\varepsilon }(1-z)^{2\Lambda +1},
\end{equation}%
\begin{equation}
y_{nl}(z)=C_{n}z^{-2\varepsilon }(1-z)^{-(2\Lambda +1)}\frac{d^{n}}{dz^{n}}%
\left[ z^{n+2\varepsilon }(1-z)^{n+2\Lambda +1}\right] .
\end{equation}%
The functions $\ y_{nl}(z)$ are, up to a numerical factor, are in the form
of\ Jacobi polynomials, i.e., $\ y_{nl}(z)\simeq P_{n}^{(2\varepsilon
,2\Lambda +1)}(1-2z),$ valid physically in the interval $(0\leq r<\infty $ $%
\rightarrow $ $0\leq z\leq 1)$ [54]. Therefore, the radial part of the wave
functions can be found by substituting Eqs. (28) and (30) into $%
R_{nl}(z)=\phi (z)y_{nl}(z)$ as%
\begin{equation}
R_{nl}(z)=N_{nl}z^{\varepsilon }(1-z)^{1+\Lambda }P_{n}^{(2\varepsilon
,2\Lambda +1)}(1-2z),
\end{equation}%
where $\varepsilon $ and $\Lambda $ are given in Eqs. (14) and (19) and $%
N_{nl}$ is a normalization constant. This equation satisfies the
requirements; $R_{nl}(z)=0$ as $z=0$ $(r\rightarrow \infty )$ and $%
R_{nl}(z)=0$ as $z=1$ $(r=0).$ Therefore, the wave functions, $R_{nl}(z)$ in
Eq. (31) is valid physically in the closed interval $z\in \lbrack 0,1]$ or $%
r\in (0,\infty ).$ Further, the wave functions satisfy the normalization
condition%
\begin{equation}
\int\limits_{0}^{\infty }\left\vert R_{nl}(r)\right\vert
^{2}dr=1=b\int\limits_{0}^{1}z^{-1}\left\vert R_{nl}(z)\right\vert ^{2}dz,
\end{equation}%
where $N_{nl}$ can be determined via%
\begin{equation}
1=bN_{nl}^{2}\int\limits_{0}^{1}z^{2\varepsilon -1}(1-z)^{2\Lambda +2}\left[
P_{n}^{(2\varepsilon ,2\Lambda +1)}(1-2z)\right] ^{2}dz.
\end{equation}%
The Jacobi polynomials, $P_{n}^{(\rho ,\nu )}(\xi ),$ can be explicitly
written in two different ways [55,56]:%
\begin{equation}
P_{n}^{(\rho ,\nu )}(\xi )=2^{-n}\sum\limits_{p=0}^{n}(-1)^{n-p}\binom{%
n+\rho }{p}\binom{n+\nu }{n-p}\left( 1-\xi \right) ^{n-p}\left( 1+\xi
\right) ^{p},
\end{equation}

\begin{equation}
P_{n}^{(\rho ,\nu )}(\xi )=\frac{\Gamma (n+\rho +1)}{n!\Gamma (n+\rho +\nu
+1)}\sum\limits_{r=0}^{n}\binom{n}{r}\frac{\Gamma (n+\rho +\nu +r+1)}{\Gamma
(r+\rho +1)}\left( \frac{\xi -1}{2}\right) ^{r},
\end{equation}%
where $\binom{n}{r}=\frac{n!}{r!(n-r)!}=\frac{\Gamma (n+1)}{\Gamma
(r+1)\Gamma (n-r+1)}.$ Using Eqs. (34)-(35), we obtain the explicit
expressions for $P_{n}^{(2\varepsilon ,2\Lambda +1)}(1-2z):$%
\[
P_{n}^{(2\varepsilon ,2\Lambda +1)}(1-2z)=(-1)^{n}\Gamma (n+2\varepsilon
+1)\Gamma (n+2\Lambda +2) 
\]

\begin{equation}
\times \sum\limits_{p=0}^{n}\frac{(-1)^{p}}{p!(n-p)!\Gamma (p+2\Lambda
+2)\Gamma (n+2\varepsilon -p+1)}z^{n-p}(1-z)^{p},
\end{equation}

\begin{equation}
P_{n}^{(2\varepsilon ,2\Lambda +1)}(1-2z)=\frac{\Gamma (n+2\varepsilon +1)}{%
\Gamma (n+2\varepsilon +2\Lambda +2)}\sum\limits_{r=0}^{n}\frac{%
(-1)^{r}\Gamma (n+2\varepsilon +2\Lambda +r+2)}{r!(n-r)!\Gamma (2\varepsilon
+r+1)}z^{r}.
\end{equation}%
Inserting Eqs. (36)-(37) into Eq. (33), one obtains%
\[
1=bN_{nl}^{2}(-1)^{n}\frac{\Gamma (n+2\Lambda +2)\Gamma (n+2\varepsilon
+1)^{2}}{\Gamma (n+2\varepsilon +2\Lambda +2)} 
\]%
\begin{equation}
\times \sum\limits_{p,r=0}^{n}\frac{(-1)^{p+r}\Gamma (n+2\varepsilon
+2\Lambda +r+2)}{p!r!(n-p)!(n-r)!\Gamma (p+2\Lambda +2)\Gamma
(n+2\varepsilon -p+1)\Gamma (2\varepsilon +r+1)}I_{nl}(p,r),
\end{equation}%
where%
\begin{equation}
I_{nl}(p,r)=\int\limits_{0}^{1}z^{n+2\varepsilon +r-p-1}(1-z)^{p+2\Lambda
+2}dz.
\end{equation}%
Using the following integral representation of the hypergeometric function
[55.56]%
\[
_{2}F_{1}(\alpha _{0},\beta _{0}:\gamma _{0};1)\frac{\Gamma (\alpha
_{0})\Gamma (\gamma _{0}-\alpha _{0})}{\Gamma (\gamma _{0})}%
=\int\limits_{0}^{1}z^{\alpha _{0}-1}(1-z)^{\gamma _{0}-\alpha
_{0}-1}(1-z)^{-\beta _{0}}dz, 
\]

\begin{equation}
\func{Re}(\gamma _{0})>\func{Re}(\alpha _{0})>0,
\end{equation}%
which gives%
\begin{equation}
_{2}F_{1}(\alpha _{0},\beta _{0}:\alpha _{0}+1;1)/\alpha
_{0}=\int\limits_{0}^{1}z^{\alpha _{0}-1}(1-z)^{-\beta _{0}}dz,
\end{equation}%
where%
\[
_{2}F_{1}(\alpha _{0},\beta _{0}:\gamma _{0};1)=\frac{\Gamma (\gamma
_{0})\Gamma (\gamma _{0}-\alpha _{0}-\beta _{0})}{\Gamma (\gamma _{0}-\alpha
_{0})\Gamma (\gamma _{0}-\beta _{0})}, 
\]%
\begin{equation}
(\func{Re}(\gamma _{0}-\alpha _{0}-\beta _{0})>0,\text{ }\func{Re}(\gamma
_{0})>\func{Re}(\beta _{0})>0).
\end{equation}%
For the present case, with the aid of Eq. (40), when $\alpha
_{0}=n+2\varepsilon +r-p,$ $\beta _{0}=-p-2\Lambda -2,$ and $\gamma
_{0}=\alpha _{0}+1$ are substituted into Eq. (41)$,$ we obtain%
\begin{equation}
I_{nl}(p,r)=\frac{_{2}F_{1}(\alpha _{0},\beta _{0}:\gamma _{0};1)}{\alpha
_{0}}=\frac{\Gamma (n+2\varepsilon +r-p+1)\Gamma (p+2\Lambda +3)}{%
(n+2\varepsilon +r-p)\Gamma (n+2\varepsilon +r+2\Lambda +3)}.
\end{equation}%
Finally, we obtain%
\[
1=bN_{nl}^{2}(-1)^{n}\frac{\Gamma (n+2\Lambda +2)\Gamma (n+2\varepsilon
+1)^{2}}{\Gamma (n+2\varepsilon +2\Lambda +2)} 
\]%
\begin{equation}
\times \sum\limits_{p,r=0}^{n}\frac{(-1)^{p+r}\Gamma (n+2\varepsilon
+r-p+1)(p+2\Lambda +2)}{p!r!(n-p)!(n-r)!\Gamma (n+2\varepsilon -p+1)\Gamma
(2\varepsilon +r+1)(n+2\varepsilon +r+2\Lambda +2)},
\end{equation}%
which gives%
\begin{equation}
N_{nl}=\frac{1}{\sqrt{s(n)}},
\end{equation}%
where%
\[
s(n)=b(-1)^{n}\frac{\Gamma (n+2\Lambda +2)\Gamma (n+2\varepsilon +1)^{2}}{%
\Gamma (n+2\varepsilon +2\Lambda +2)} 
\]%
\begin{equation}
\times \sum\limits_{p,r=0}^{n}\frac{(-1)^{p+r}\Gamma (n+2\varepsilon
+r-p+1)(p+2\Lambda +2)}{p!r!(n-p)!(n-r)!\Gamma (n+2\varepsilon -p+1)\Gamma
(2\varepsilon +r+1)(n+2\varepsilon +r+2\Lambda +2)}.
\end{equation}

\section{Numerical Results}

To show the accuracy of our results, we calculate the energy eigenvalues for
various $n$ and $l$ quantum numbers with two different values of the
parameters $\alpha .$ Its shown in Table 1, the present approximately
numerical results are not in a good agreement when long potential range
(small values of parameter $b$). The energy eigenvalues for short potential
range (large values of parameter $b$) are in agreement with the other
authors. The energy spectra for various diatomic molecules like $HCl,CH,LiH$
and $CO$ are presented in Tables 2 and 3.

\section{Discussions}

In this work, we have utilized the hypergeometric method and solved the
radial $\mathrm{SE}$ for the M-R model potential with the angular momentum $%
l\neq 0$ states$.$ We have derived the binding energy spectra in Eq. (26)
and their corresponding wave functions in Eq. (31).

Let us study special cases. We have shown that for $\alpha =0$ $(1)$, the
present solution reduces to the one of the Hulth\'{e}n potential [16,18,19]:%
\begin{equation}
V^{(H)}(r)=-V_{0}\frac{e^{-\delta r}}{1-e^{-\delta r}},\text{ }%
V_{0}=Ze^{2}\delta ,\text{ }\delta =b^{-1}
\end{equation}%
where $Ze^{2}$ is the strength and $\delta $ is the screening parameter and $%
b$ is the range of potential. If the potential is used for atoms, the $Z$ is
identified with the atomic number. This can be achieved by setting $\Lambda
=l,$ hence, the energy for $l\neq 0$ states%
\begin{equation}
E_{nl}=-\frac{\left[ A-(n+l+1)^{2}\right] ^{2}\hbar ^{2}}{8\mu
b^{2}(n+l+1)^{2}},\text{ \ }0\leq n,l<\infty .
\end{equation}%
and for $s$-wave ($l=0)$ states%
\begin{equation}
E_{n}=-\frac{\left[ A-(n+1)^{2}\right] ^{2}\hbar ^{2}}{8\mu b^{2}(n+1)^{2}},%
\text{ \ }0\leq n<\infty
\end{equation}%
Essentially, these results coincide with those obtained by the Feynman
integral method [31] and the standard way [32,33], respectively.
Furthermore, if taking $b=1/\delta $ and identifying $\frac{A\hbar ^{2}}{%
2\mu b^{2}}$ as $Ze^{2}\delta ,$ we are able to obtain%
\begin{equation}
E_{nl}=-\frac{\mu \left( Ze^{2}\right) ^{2}}{2\hbar ^{2}}\left[ \frac{1}{%
n+l+1}-\frac{\hbar ^{2}\delta }{2Ze^{2}\mu }(n+l+1)\right] ^{2},
\end{equation}%
which coincides with those of Refs. [16,18]. With natural units $\hbar
^{2}=\mu =Z=e=1,$ we have%
\begin{equation}
E_{nl}=-\frac{1}{2}\left[ \frac{1}{n+l+1}-\frac{(n+l+1)}{2}\delta \right]
^{2},
\end{equation}%
which coincides with Refs. [16,33].

The corresponding radial wave functions are expressed as%
\begin{equation}
R_{nl}(r)=N_{nl}e^{-\delta \varepsilon r}(1-e^{-\delta
r})^{l+1}P_{n}^{(2\varepsilon ,2l+1)}(1-2e^{-\delta r}),
\end{equation}%
where%
\begin{equation}
\varepsilon =\frac{\mu Ze^{2}}{\hbar ^{2}\delta }\left[ \frac{1}{n+l+1}-%
\frac{\hbar ^{2}\delta }{2Ze^{2}\mu }(n+l+1)\right] ,\text{ }0\leq
n,l<\infty ,
\end{equation}%
which coincides for the ground state with that given in Eq. (6) by G\"{o}n%
\"{u}l \textit{et al.} [18]. In addition, for $\delta r\ll 1$ (i.e., $r/b\ll
1),$ the Hulth\'{e}n potential turns to become a Coulomb potential: $%
V(r)=-Ze^{2}/r$ with energy levels and wavefunctions:%
\[
E_{nl}=-\frac{\varepsilon _{0}}{(n+l+1)^{2}},\text{ }n=0,1,2,.. 
\]

\begin{equation}
\varepsilon _{0}=\frac{Z^{2}\hbar ^{2}}{2\mu a_{0}^{2}},\text{ }a_{0}=\frac{%
\hbar ^{2}}{\mu e^{2}}
\end{equation}%
where $\varepsilon _{0}=13.6$ $eV$ and $a_{0}$ is Bohr radius for the
Hydrogen atom. The wave functions are%
\[
R_{nl}=N_{nl}\exp \left[ -\frac{\mu Ze^{2}}{\hbar ^{2}}\frac{r}{\left(
n+l+1\right) }\right] r^{l+1}P_{n}^{\left( \frac{2\mu Ze^{2}}{\hbar
^{2}\delta (n+l+1)},2l+1\right) }(1+2\delta r) 
\]%
which coincide with Refs. [3,16,22].

\section{Cocluding Remarks}

In this work, we have presented the approximate solutions of the $l$-wave
Schr\"{o}dinger equation with the M-R potential. The special cases for $%
\alpha =0,1$ are discussed. The results are in good agreement with those
obtained by other methods for short potential range, small $\alpha $ and $l.$
We have also studied two special cases for $l=0,$ $l\neq 0$ and Hulth\'{e}n
potential. The results we have ended up show that the N-U method constitute
a reliable alternative way in solving the exponential potentials.

\acknowledgments The support provided by the Scientific and Technological
Research Council of Turkey (T\"{U}B\.{I}TAK) is highly appreciated.

\bigskip

\bigskip \baselineskip= 2\baselineskip
\bigskip

\begin{table}[tbp]
\caption{Eigenvalues ($-E_{nl}$) for $2p,3p,3d,4p,4d,4f,5p,5d,5f,5g,6p,6d,6f$
and $6g$ states in atomic units ($\hbar =\protect\mu =1)$ and for $\protect%
\alpha =0.75$ and $\protect\alpha =1.5,$ $A=2b.$}%
\begin{tabular}{llllllll}
&  & $\alpha =0.75$ &  &  & $\alpha =1.5$ &  &  \\ 
states & $1/b$ & present & QD [33] & LS [57] & present & QD [33] & LS [57]
\\ 
\tableline$2p$ & $0.025$ & $0.1205793$ & $0.1205793$ & $0.1205271$ & $%
0.0900228$ & $0.0900229$ & $0.0899708$ \\ 
& $0.050$ & $0.1084228$ & $0.1084228$ & $0.1082151$ & $0.0802472$ & $%
0.0802472$ & $0.0800400$ \\ 
& $0.075$ & $0.0969120$ & $0.0969120$ & $0.0964469$ & $0.0710332$ & $%
0.0710332$ & $0.0705701$ \\ 
& $0.100$ & $0.0860740$ &  &  & $0.0577157$ &  &  \\ 
$3p$ & $0.025$ & $0.0459296$ & $0.0459297$ & $0.0458779$ & $0.0369650$ & $%
0.0369651$ & $0.0369134$ \\ 
& $0.050$ & $0.0352672$ & $0.0352672$ & $0.0350633$ & $0.0274719$ & $%
0.0274719$ & $0.0272696$ \\ 
& $0.075$ & $0.0260109$ & $0.0260110$ & $0.0255654$ & $0.0193850$ & $%
0.0193850$ & $0.0189474$ \\ 
& $0.100$ & $0.0181609$ &  &  & $0.0127043$ &  &  \\ 
$3d$ & $0.025$ & $0.0449299$ & $0.0449299$ & $0.0447743$ & $0.0396344$ & $%
0.0396345$ & $0.0394789$ \\ 
& $0.050$ & $0.0343082$ & $0.0343082$ & $0.0336930$ & $0.0300629$ & $%
0.0300629$ & $0.0294496$ \\ 
& $0.075$ & $0.0251168$ & $0.0251168$ & $0.0237621$ & $0.0218120$ & $%
0.0218121$ & $0.0204663$ \\ 
$4p$ & $0.025$ & $0.0208608$ & $0.0208608$ & $0.0208097$ & $0.0172249$ & $%
0.0172249$ & $0.0171740$ \\ 
& $0.050$ & $0.0119291$ & $0.0119292$ & $0.0117365$ & $0.0091019$ & $%
0.0091019$ & $0.0089134$ \\ 
& $0.075$ & $0.0054773$ & $0.0054773$ & $0.0050945$ & $0.0035478$ & $%
0.0035478$ & $0.0031884$ \\ 
$4d$ & $0.025$ & $0.0204555$ & $0.0204555$ & $0.0203017$ & $0.0183649$ & $%
0.0183649$ & $0.0182115$ \\ 
& $0.050$ & $0.0115741$ & $0.0115742$ & $0.0109904$ & $0.0100947$ & $%
0.0100947$ & $0.0095167$ \\ 
& $0.075$ & $0.0052047$ & $0.0052047$ & $0.0040331$ & $0.0042808$ & $%
0.0042808$ & $0.0031399$ \\ 
$4f$ & $0.025$ & $0.0202886$ & $0.0202887$ & $0.0199797$ & $0.0189222$ & $%
0.0189223$ & $0.0186137$ \\ 
& $0.050$ & $0.0114283$ & $0.0114284$ & $0.0102393$ & $0.0105852$ & $%
0.0105852$ & $0.0094015$ \\ 
& $0.075$ & $0.0050935$ & $0.0050935$ & $0.0026443$ & $0.0046527$ & $%
0.0046527$ & $0.0022307$ \\ 
$5p$ & $0.025$ & $0.0098576$ & $0.0098576$ & $0.0098079$ & $0.0081308$ & $%
0.0081308$ & $0.0080816$ \\ 
$5d$ & $0.025$ & $0.0096637$ & $0.0096637$ & $0.0095141$ & $0.0086902$ & $%
0.0086902$ & $0.0085415$ \\ 
$5f$ & $0.025$ & $0.0095837$ & $0.0095837$ & $0.0092825$ & $0.0089622$ & $%
0.0089622$ & $0.0086619$ \\ 
$5g$ & $0.025$ & $0.0095398$ & $0.0095398$ & $0.0090330$ & $0.0091210$ & $%
0.0091210$ & $0.0086150$ \\ 
$6p$ & $0.025$ & $0.0044051$ & $0.0044051$ & $0.0043583$ & $0.0035334$ & $%
0.0035334$ & $0.0034876$ \\ 
$6d$ & $0.025$ & $0.0043061$ & $0.0043061$ & $0.0041650$ & $0.0038209$ & $%
0.0038209$ & $0.0036813$ \\ 
$6f$ & $0.025$ & $0.0042652$ & $0.0042652$ & $0.0039803$ & $0.0039606$ & $%
0.0039606$ & $0.0036774$ \\ 
$6g$ & $0.025$ & $0.0042428$ & $0.0042428$ & $0.0037611$ & $0.0040422$ & $%
0.0040422$ & $0.0035623$%
\end{tabular}%
\end{table}

\begin{table}[tbp]
\caption{Eigenvalues ($-E_{nl}$) of $HCl$ and $CH$ (in $eV$) for $%
2p,3p,3d,4p,4d,4f,5p,5d,5f,$ $5g,6p,6d,6f$ and $6g$ states where $\hbar
c=1973.29$ $eV$ $A^{\circ },$ $\protect\mu _{HCl}=0.9801045$ $amu,$ $\protect%
\mu _{CH}=0.929931$ $amu$ and $A=2b.$}%
\begin{tabular}{llllllll}
states & $1/b$\tablenotetext[1]{$b$ is in $pm$.}\tablenotemark[1] & $HCl/$ $%
\alpha =0,1$ & $\alpha =0.75$ & $\alpha =1.5$ & $CH/$ $\alpha =0,1$ & $%
\alpha =0.75$ & $\alpha =1.5$ \\ 
\tableline$2p$ & $0.025$ & $4.81152646$ & $5.14278553$ & $3.83953094$ & $%
5.07112758$ & $5.42025940$ & $4.04668901$ \\ 
& $0.050$ & $4.31837832$ & $4.62430290$ & $3.42259525$ & $4.55137212$ & $%
4.87380256$ & $3.60725796$ \\ 
& $0.075$ & $3.85188684$ & $4.13335980$ & $3.02961216$ & $4.05971155$ & $%
4.35637111$ & $3.19307186$ \\ 
& $0.100$ & $3.41205201$ & $3.66996049$ & $2.46161213$ & $3.59614587$ & $%
3.86796955$ & $2.59442595$ \\ 
$3p$ & $0.025$ & $1.86633700$ & $1.95892730$ & $1.57658128$ & $1.96703335$ & 
$2.06461927$ & $1.66164415$ \\ 
& $0.050$ & $1.42316902$ & $1.50416901$ & $1.17169439$ & $1.49995469$ & $%
1.58532495$ & $1.23491200$ \\ 
& $0.075$ & $1.03998066$ & $1.10938179$ & $0.82678285$ & $1.09609178$ & $%
1.16923738$ & $0.87139110$ \\ 
& $0.100$ & $0.71676763$ & $0.77457419$ & $0.54184665$ & $0.75544012$ & $%
0.81636557$ & $0.57108145$ \\ 
$3d$ & $0.025$ & $1.86633700$ & $1.91628944$ & $1.69043293$ & $1.96703335$ & 
$2.01968093$ & $1.78163855$ \\ 
& $0.050$ & $1.42316902$ & $1.46326703$ & $1.28220223$ & $1.49995469$ & $%
1.54221615$ & $1.35138217$ \\ 
& $0.075$ & $1.03998066$ & $1.07124785$ & $0.93029598$ & $1.09609178$ & $%
1.12904596$ & $0.98048917$ \\ 
& $0.100$ & $0.71676763$ & $0.74022762$ & $0.63472271$ & $0.75544012$ & $%
0.78016587$ & $0.66896854$ \\ 
$4p$ & $0.025$ & $0.85301300$ & $0.88972668$ & $0.73465318$ & $0.89903647$ & 
$0.93773100$ & $0.77429066$ \\ 
& $0.050$ & $0.47981981$ & $0.50878387$ & $0.38820195$ & $0.50570801$ & $%
0.53623480$ & $0.40914700$ \\ 
& $0.075$ & $0.21325325$ & $0.23361041$ & $0.15131598$ & $0.22475912$ & $%
0.24621462$ & $0.15948008$ \\ 
$4d$ & $0.025$ & $0.85301300$ & $0.87244037$ & $0.78327492$ & $0.89903647$ & 
$0.91951202$ & $0.82553574$ \\ 
& $0.050$ & $0.47981981$ & $0.49364289$ & $0.43054552$ & $0.50570801$ & $%
0.52027690$ & $0.45377517$ \\ 
& $0.075$ & $0.21325325$ & $0.22198384$ & $0.18257890$ & $0.22475912$ & $%
0.23396076$ & $0.19242977$ \\ 
$4f$ & $0.025$ & $0.85301300$ & $0.86532198$ & $0.80704413$ & $0.89903647$ & 
$0.91200956$ & $0.85058739$ \\ 
& $0.050$ & $0.47981981$ & $0.48742442$ & $0.45146566$ & $0.50570801$ & $%
0.51372292$ & $0.47582404$ \\ 
& $0.075$ & $0.21325325$ & $0.21724109$ & $0.19844068$ & $0.22475912$ & $%
0.22896211$ & $0.20914735$ \\ 
$5p$ & $0.025$ & $0.40318193$ & $0.42043305$ & $0.34678391$ & $0.42493521$ & 
$0.44311709$ & $0.36549429$ \\ 
$5d$ & $0.025$ & $0.40318193$ & $0.41216309$ & $0.37064268$ & $0.42493521$ & 
$0.43440094$ & $0.39064034$ \\ 
$5f$ & $0.025$ & $0.40318193$ & $0.40875104$ & $0.38224366$ & $0.42493521$ & 
$0.43080479$ & $0.40286723$ \\ 
$5g$ & $0.025$ & $0.40318193$ & $0.40687867$ & $0.38901658$ & $0.42493521$ & 
$0.42883140$ & $0.41000558$ \\ 
$6p$ & $0.025$ & $0.17919244$ & $0.18788038$ & $0.15070181$ & $0.18886059$ & 
$0.19801728$ & $0.15883277$ \\ 
$6d$ & $0.025$ & $0.17919244$ & $0.18365796$ & $0.16296387$ & $0.18886059$ & 
$0.19356705$ & $0.17175642$ \\ 
$6f$ & $0.025$ & $0.17919244$ & $0.18191355$ & $0.16892216$ & $0.18886059$ & 
$0.19172852$ & $0.17803620$ \\ 
$6g$ & $0.025$ & $0.17919244$ & $0.18095818$ & $0.17240246$ & $0.18886059$ & 
$0.19072160$ & $0.18170426$%
\end{tabular}%
\end{table}

\begin{table}[tbp]
\caption{Eigenvalues ($-E_{nl}$) of $LiH$ and $CO$ (in $eV$) for $%
2p,3p,3d,4p,4d,4f,5p,5d,$ $5f,5g,6p,6d,6f$ and $6g$ states where $\hbar
c=1973.29$ $eV$ $A^{\circ },$ $\protect\mu _{LiH}=0.8801221$ $amu,$ $\protect%
\mu _{CO}=6.8606719$ $amu$ and $A=2b.$}%
\begin{tabular}{llllllll}
states & $1/b$\tablenotemark[1]\tablenotetext[1]{$b$ is in $pm$.} & $LiH/$ $%
\alpha =0,1$ & $\alpha =0.75$ & $\alpha =1.5$ & $CO/$ $\alpha =0,1$ & $%
\alpha =0.75$ & $\alpha =1.5$ \\ 
\tableline$2p$ & $0.025$ & $5.35811876$ & $5.72700906$ & $4.27570397$ & $%
1.374733789$ & $0.734690030$ & $0.548509185$ \\ 
& $0.050$ & $4.80894870$ & $5.14962650$ & $3.81140413$ & $1.233833096$ & $%
0.660620439$ & $0.488946426$ \\ 
& $0.075$ & $4.28946350$ & $4.60291196$ & $3.37377792$ & $1.100548657$ & $%
0.590485101$ & $0.432805497$ \\ 
& $0.100$ & $3.79966317$ & $4.08687021$ & $2.74125274$ & $0.974880471$ & $%
0.524284624$ & $0.351661930$ \\ 
$3p$ & $0.025$ & $2.07835401$ & $2.18146262$ & $1.75568186$ & $0.533243776$
& $0.279849188$ & $0.225227854$ \\ 
& $0.050$ & $1.58484188$ & $1.67504351$ & $1.30479958$ & $0.406623254$ & $%
0.214883153$ & $0.167386368$ \\ 
& $0.075$ & $1.15812308$ & $1.23540823$ & $0.92070588$ & $0.297139912$ & $%
0.158484490$ & $0.118112862$ \\ 
& $0.100$ & $0.79819287$ & $0.86256629$ & $0.60340076$ & $0.204792531$ & $%
0.110654417$ & $0.077407337$ \\ 
$3d$ & $0.025$ & $2.07835401$ & $2.13398108$ & $1.88246712$ & $0.533243776$
& $0.273758013$ & $0.241492516$ \\ 
& $0.050$ & $1.58484188$ & $1.62949505$ & $1.42786117$ & $0.406623254$ & $%
0.209039964$ & $0.183173338$ \\ 
& $0.075$ & $1.15812308$ & $1.19294225$ & $1.03597816$ & $0.299139912$ & $%
0.153036736$ & $0.132900580$ \\ 
& $0.100$ & $0.79819287$ & $0.82431793$ & $0.70682759$ & $0.204792531$ & $%
0.105747722$ & $0.090675460$ \\ 
$4p$ & $0.025$ & $0.94991579$ & $0.99080017$ & $0.81811023$ & $0.243720118$
& $0.127104916$ & $0.104951366$ \\ 
& $0.050$ & $0.53432763$ & $0.56658202$ & $0.43230193$ & $0.137092566$ & $%
0.072684041$ & $0.055457903$ \\ 
& $0.075$ & $0.23747895$ & $0.26014869$ & $0.16850556$ & $0.060930029$ & $%
0.033373205$ & $0.021616756$ \\ 
$4d$ & $0.025$ & $0.94991579$ & $0.97155012$ & $0.87225543$ & $0.243720118$
& $0.124635422$ & $0.111897390$ \\ 
& $0.050$ & $0.53432763$ & $0.54972102$ & $0.47945575$ & $0.137092566$ & $%
0.070521025$ & $0.061507037$ \\ 
& $0.075$ & $0.23747895$ & $0.24720134$ & $0.20331998$ & $0.060930029$ & $%
0.031712252$ & $0.026082927$ \\ 
$4f$ & $0.025$ & $0.94991579$ & $0.96362308$ & $0.89872483$ & $0.243720118$
& $0.123618500$ & $0.115293020$ \\ 
& $0.050$ & $0.53432763$ & $0.54279613$ & $0.50275243$ & $0.137092566$ & $%
0.069632666$ & $0.064495655$ \\ 
& $0.075$ & $0.23747895$ & $0.24191980$ & $0.22098366$ & $0.060930029$ & $%
0.031034710$ & $0.028348915$ \\ 
$5p$ & $0.025$ & $0.44898364$ & $0.46819450$ & $0.38617877$ & $0.115195837$
& $0.060062386$ & $0.049540988$ \\ 
$5d$ & $0.025$ & $0.44898364$ & $0.45898506$ & $0.41274791$ & $0.115195837$
& $0.058880953$ & $0.052949414$ \\ 
$5f$ & $0.025$ & $0.44898364$ & $0.45518540$ & $0.42566677$ & $0.115195837$
& $0.058393512$ & $0.054606711$ \\ 
$5g$ & $0.025$ & $0.44898364$ & $0.45310033$ & $0.43320910$ & $0.115195837$
& $0.058126029$ & $0.055574280$ \\ 
$6p$ & $0.025$ & $0.19954881$ & $0.20922370$ & $0.16782162$ & $0.051198285$
& $0.026840287$ & $0.021529017$ \\ 
$6d$ & $0.025$ & $0.19954881$ & $0.20452162$ & $0.18147666$ & $0.051198285$
& $0.026237080$ & $0.023280755$ \\ 
$6f$ & $0.025$ & $0.19954881$ & $0.20257904$ & $0.18811182$ & $0.051198285$
& $0.025987876$ & $0.024131947$ \\ 
$6g$ & $0.025$ & $0.19954881$ & $0.20151514$ & $0.19198748$ & $0.051198285$
& $0.025851393$ & $0.024629136$%
\end{tabular}%
\end{table}

\end{document}